\documentclass[twocolumn,prl,superscriptaddress,floatfix,showpacs]{revtex4}
\usepackage{bm}
\usepackage{epsfig}

\def\URuSi{URu$_2$Si$_2$}

\def\urx{U(Ru$_{0.98}$Rh$_{0.02}$)$_2$Si$_2$}
\def\urs{URu$_2$Si$_2$}

\def\jpcm{J.\ Phys.\ Cond.\ Matter\ }
\def\jmmm{J.\ Magn.\ Magn.\ Mater.\ }

\def\pb{Physica B\ }

\begin{document}

\title{Pressure dependence of magnetism in \URuSi}

\author{F. Bourdarot}
\author{B. Fak}
\author{V.P. Mineev}
\author{M.E. Zhitomirsky}
\author{N. Kernavanois}
\author{S. Raymond}
\author{P. Burlet}
\affiliation{DRFMC, SPSMS, CEA Grenoble, 38054 Grenoble, France}
\author{F. Lapierre}
\author{P. Lejay}
\affiliation{Centre de Recherches sur les Tres Basses Temperatures,  CNRS, BP 166, 38042 Grenoble, France}
\author{J. Flouquet}
\affiliation{DRFMC, SPSMS, CEA Grenoble, 38054 Grenoble, France}

\date{29 August 2003; printed \today}

\begin{abstract}
Neutron-scattering and specific-heat measurements 
of the heavy-fermion superconductor \urs\
under hydrostatic pressure and with Rh-doping [\urx]
show the existence of two magnetic phase transitions. 
At the second-order phase transition $T_m\!\sim$17.5 K, 
a tiny ordered moment is established, while 
at $T_M\!<\!T_m$, a first-order phase transition (under pressure or doping) 
gives rise to a large moment. The results can be understood in terms of 
a hidden OP $\psi$ coupled to the ordered moment $m$,
where $m$ and $\psi$ have the same symmetry.
 \end{abstract}
\pacs{75.25.+z, 75.30.Kz, 75.30.Cr, 75.50.Ee, 74.70.Tx} 
\maketitle

The phase transition at $T_m$=17.5 K in \urs\
is one of the most puzzling features in heavy-fermion systems.
Most bulk properties display anomalies at $T_m$, 
and in particular there is a huge jump of $\Delta C/T_m$=0.3 JK$^{-2}$mol$^{-1}$
in the specific heat \cite{palstra}. 
These anomalies cannot be explained by the tiny 
ordered 
moment of $\sim$0.03 $\mu_B$, 
which also develops at $T_m$. 
The antiferromagnetic (AFM) structure, 
observed both by neutron \cite{broholm,walker,fak,mason,santini,FRL} 
and x-ray \cite{isaacs} scattering, 
is characterized by a propagation vector  {\bf k}=(001) 
and dipolar moments along the $c$ axis 
 of the body-centered tetragonal structure (I4/mmm). 

In a recent {\it tour de force} experiment combining neutron elastic 
scattering with high magnetic fields \cite{FRL}, it was found that the magnetic 
moment in \urs\ did not disappear at 15 T as suggested by the 
extrapolation of earlier low-field data \cite{mason,santini}, but remains finite at 17 T 
with an inflection point near 7 T .  The particular 
field dependence of the ordered moment, $m(H)$, agrees with the 
predictions of a model containing two order parameters (OPs) 
with the same symmetry \cite{shah}, based
on the theoretical work of Ref. \cite{agterberg}. 

\begin{figure} 
\noindent
\begin{center}
\includegraphics[width=.85\columnwidth]{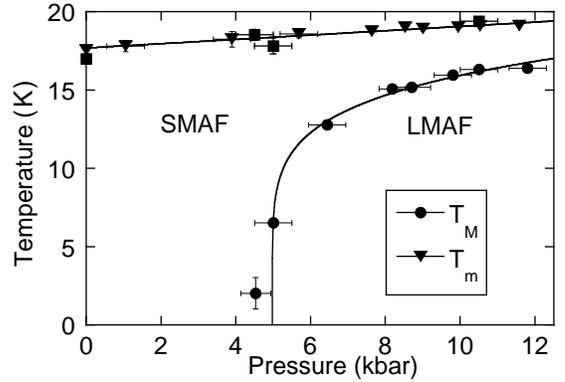}
\caption{Pressure-temperature phase diagram of \urs\ showing the SMAF and LMAF 
phases determined from neutron scattering (circles and squares)
and resistivity measurements  (triangles, Ref. \cite{schmidt}). The lines are guides to the eye.}
\label{phasediagram}
\end{center}
\end{figure}

In this Letter, we present new experimental data 
where hydrostatic or chemical pressure 
is used to tune the interactions in the system.  
The main result 
is the existence of two distinct phases in the $p-T$ phase diagram of \urs\ 
for pressures $p>p_M$, where $p_M\approx 4.9$ kbar  (see Fig.\  \ref{phasediagram}), 
or for small Rh doping.  
The existence of two phases have also been inferred from 
thermal expansion measurements \cite{sato}, but, as will be discussed later,
their phase diagram is fundamentally different from ours. 
We call the two magnetic phases for small-moment (SMAF) and 
large-moment (LMAF) antiferromagnetic phases.  
The SMAF phase, observed below the second-order phase transition 
at $T_m\approx 17.5$ K ($T_m(p)$ in Fig.\ \ref{phasediagram} 
is determined from resistivity measurements under pressure, 
Ref. \cite{schmidt}), is the same as that observed at $p$=0 
and is characterized by a small moment $m$ 
and large anomalies in the macroscopic properties.  
The low-T LMAF phase, observed below the transition $T_M$, 
which seems to be a first-order transition, 
is characterized by a large moment $m$ and small anomalies in the macroscopic properties.  
The SMAF and LMAF phases have the same magnetic structures.
 
Many phenomenological theories for \urs\ introduce two physical quantities
or ``order parameters'':
an AFM parameter $m$ that describes ordering of localized dipolar 5$f$ moments
and a hidden (unknown) order $\psi$ 
responsible for the $T_m$=17.5~K transition  \cite{shah,agterberg}. 
It has been suggested that $\psi$ could originate from 
triple-spin correlators of 5$f$ electrons  \cite{agterberg}
or an exotic spin-density wave. 
The Landau free energy for the two OP's
(up to fourth-order terms) can be written as
\begin{equation}
F=\alpha_\psi\psi^2\! +\! \alpha_m m^2\! +\! 2\gamma\psi m\! +\!
\beta_\psi \psi^4 \!+ \!\beta_m m^4 \!+\!
2\beta_i \psi^2 m^2
\label{landau}
\end{equation}
The coupling term $\psi m$ 
is allowed only if $m$ and $\psi$ 
transform according to the same irreducible representation,
otherwise $\gamma\equiv 0$.
Coupling terms of type $\psi^3m$ and $\psi m^3$ can be removed
from Eq.~(\ref{landau}) by linear transformation.
Shah {\it et al}.\ \cite{shah} studied the above
free energy  at zero pressure and finite magnetic
fields $H$ and predicted an inflection point for $m(H)$ only
in the case $\gamma\neq 0$. 
This was recently observed experimentally \cite{FRL}. 
Motivated by this agreement, we investigate here the 
functional Eq.~(\ref{landau}) as a function of 
pressure.

For $\gamma\neq 0$,  the two OP's
appear simultaneously below the transition line
from the paramagnetic state $T_m(p)$, given
by $\alpha_\psi\alpha_m=\gamma^2$.
At $p$=0, the AFM
OP behaves as $m\approx -(\gamma/\alpha_m)\psi$
(for $|\alpha_\psi|,|\gamma|\ll\alpha_m$). A small coefficient
$\gamma/\alpha_m$ implies weak ordered moments, while $\psi$
gives rise to a large anomaly in the specific heat at $T_m$,
as in the SMAF phase. 
The presence and nature
of extra transitions below $T_m(p)$ depend on quartic terms in Eq.\ (\ref{landau}). 
Since the transition at $T_M$ appears to be first order,  we consider 
the case $\beta_i > \sqrt{\beta_\psi \beta_m}$.

\begin{figure} 
\includegraphics[width=.75\columnwidth]{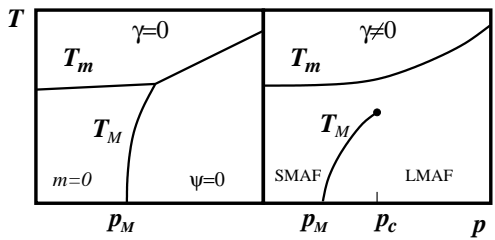}
\caption{\label{Fig1} Schematic phase  diagram for the 2OP 
functional Eq.\ (\ref{landau}) for $\gamma$=0 (left panel)
and $\gamma\!\neq\!0$ (right panel).}
\label{zhito}
\end{figure}

The $p$--$T$ phase diagram for $\gamma$=0 is sketched 
in the left panel of
Fig.~\ref{zhito}. The line of first-order phase transitions $T_M(p)$
between states with $m$=0 and $\psi$=0 emerges from
the kink point on $T_m(p)$. The equation for
$T_M(p)$ is $\alpha_\psi^2\beta_m=\alpha_m^2\beta_\psi$.
Such a phase diagram was recently considered by
Chandra {\it et al}.\ \cite{chandra} in their interpretation of small
AFM moments as a phase separation effect.

The recently observed inflection point in $m(H)$ \cite{FRL}, however,  
points to an intrinsic origin of the tiny ordered moment and a finite $\gamma$.
For $\gamma\!\neq\!0$, the first-order transition line $T_M(p)$ is stable 
and its position in the $p$--$T$ plane
is given by the same equation as for $\gamma$=0. 
However, $T_M(p)$ splits from the line of second order transitions
$T_m(p)$ and terminates at a critical point $p_c$ below
$T_m$ (right panel of Fig.\ \ref{zhito}). The location of the critical point is given by 
$\alpha_\psi^c = - 2 |\gamma|(\beta_\psi^3\beta_m)^{1/4}/
(\beta_i-\sqrt{\beta_\psi\beta_m})$. The two
states above and below $T_M(p)$ are phases with large $\psi_L$ and
small $m_S$ (SMAF) and with small $\psi_S$ and large $m_L$ (LMAF),
respectively. A relative jump of the ordered AFM
moments across $T_M(p)$ is given by $(m_L-m_S)/(m_L+m_S) =
\sqrt{(\alpha_\psi-\alpha_\psi^c)/(\alpha_\psi+\alpha_\psi^c)}$.
The size of the jump varies continuously along $T_M(p)$ and vanishes at
$p=p_c$. The corresponding phase diagram, 
shown schematically in the right panel of Fig.~\ref{zhito},
resembles the vapor-liquid phase diagram.

\begin{figure} 
\noindent
\begin{center}
\includegraphics[width=.9\columnwidth]{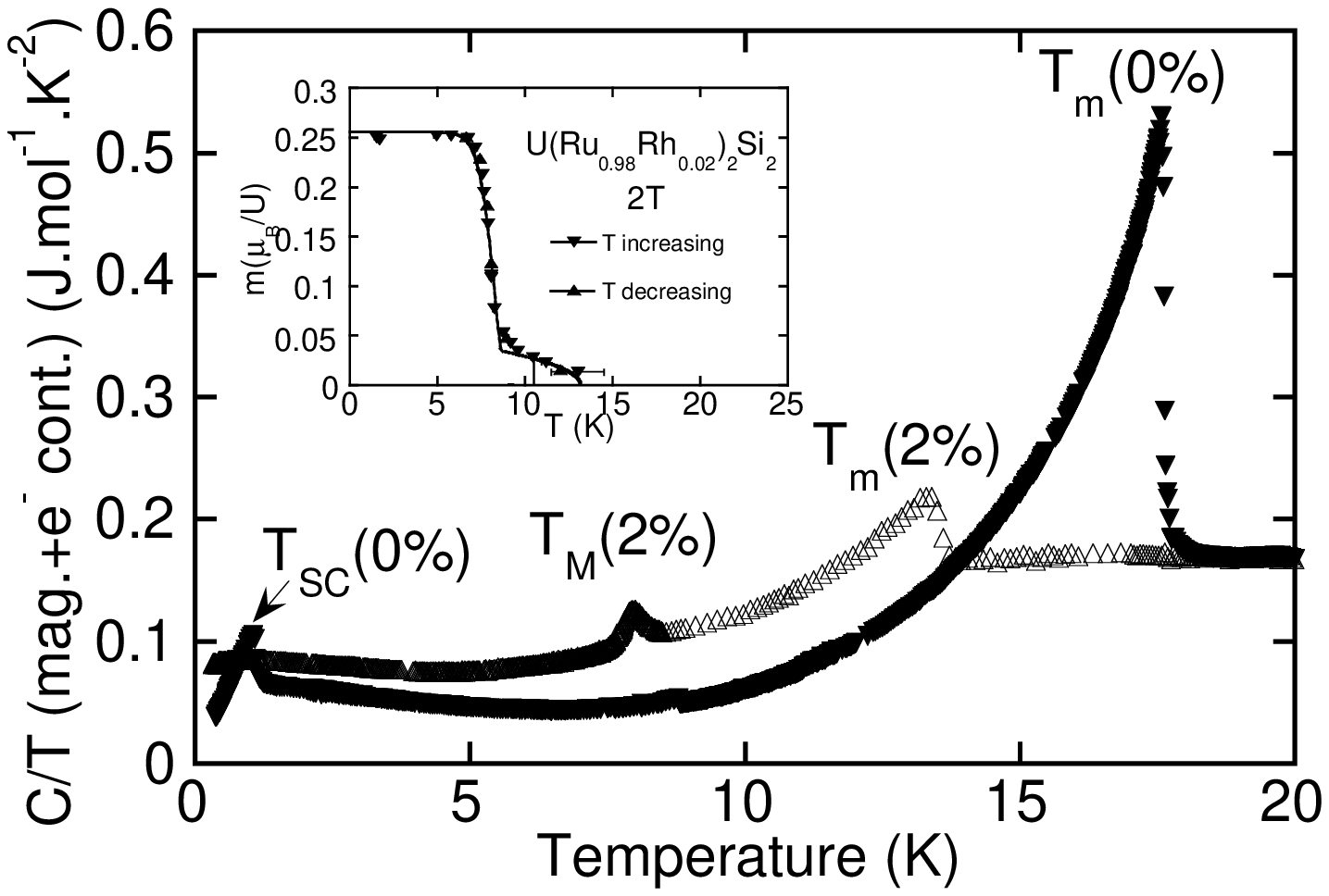}
\caption{
Electronic specific heat (after subtraction of the phonon contribution 
obtained from ThRu$_2$Si$_2$) divided by temperature for \urs\ and for \urx. 
The inset shows the $T$ dependence of the magnetic moment in  \urx\ 
for decreasing and increasing $T$.}
\label{twopercent}
\end{center}
\end{figure}
The first evidence for the existence of two phase transitions 
were obtained from neutron scattering 
and specific-heat measurements on a Rh doped sample: 
\urx\,  \cite{these}. 
The $T$ dependence of the ordered magnetic moment  
(see inset of Fig.\  \ref{twopercent}) clearly shows two transitions at 
$T_M$=8.3 K 
and $T_m$=13.2 K, with saturation moments (extrapolated to $T\!\rightarrow\!0$) of 0.25 and 0.04
$\mu_B$, respectively.  
 The transition at $T_M$ has a first order shape (although no hysteresis was observed) 
 while that at $T_m$ is second order.
These measurements were performed on the DN3 neutron diffractometer at the Silo\'e reactor 
using a wavelength of 1.54 \AA\ 
and on the thermal triple-axis spectrometer 2T at the Orph\'ee reactor 
with a final wave vector {\bf $k_f $}= 2.662 \AA$^{-1}$ 
on a 1.2 g annealed Czochralski-grown single crystal of \urx. 

Specific heat measurements  (see Fig.\  \ref{twopercent})
show that while 
the large jump in the specific heat is associated with a small moment,
the small jump is associated with a large moment estimated at 0.16 $\mu_B$ 
 \cite{calcp}. 
This shows that while there is no relation between the small moment $m_S$ and $\Delta C/T_m$, 
the large moment $m_L$  is consistent with the specific heat anomaly at $T_M$. 
The entropy  jump of the LMAF transition deduced from these measurements corresponds to 
$\Delta S(T_M)$ = 23 mJK$^{-1}$mol$^{-1}$. 
The effect of Rh doping is more  subtle than a simple (negative) pressure, 
since Rh is not isoelectronic with Ru. 
The added electron  modifies the band structure, and while the volume increases slightly, 
the low-$T$ lattice parameter $a$ decreases and $c$ increases, 
as shown by X-ray measurements  on the ID31 beam line at the ESRF \cite{id20}. 
Using the Clausius-Clapeyron relation with $\Delta V$ deduced from the $pure$ \urs \cite{sato}, 
a large slope $dT_M/dp$ is found, characteristic of a first-order phase transition.

\begin{figure} 
\includegraphics[width=.9\columnwidth,clip]{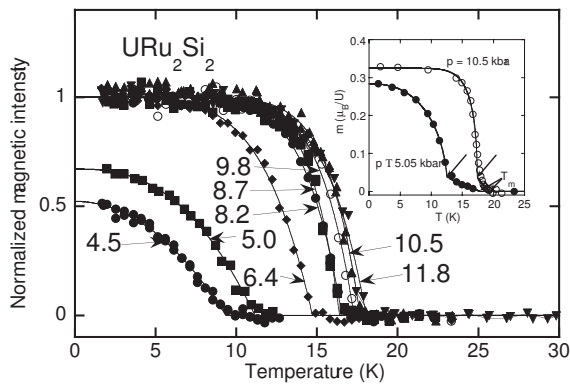}
\noindent
\begin{center}
\caption{Normalized intensity ($\propto m^2$) of the ordered magnetic moment in \URuSi\
as a function of  increasing temperature for different pressures. 
The pressures 8.7, 10.5, and 11.8 were also measured for decreasing temperature. 
The inset shows the temperature dependence of the magnetic moment $m$ 
at $p$=5.05 kbar and 10.5 kbar with the two transitions. The arrows show the onset of the LMAF phase.}
\label{MvsTpress}
\end{center}
\end{figure}
Experiments under hydrostatic pressure on the pure compound 
show strong similarities with the Rh-doped sample. 
Neutron-scattering measurements were performed at the Institut Laue-Langevin (ILL) 
on the D15 diffractometer using a wavelength of 0.85 \AA\ 
and on the IN22 triple-axis spectrometer using a final wave vector of 2.662 \AA$^{-1}$. 
A 4.8-mm diameter crystal was cut from the same batch used in the high-field measurements of Ref.\  \cite{FRL} 
with the $a$ axis along the 4-mm long cylinder axis. 
Standard ILL CuBe and steel clamp pressure cells of inner diameter of 5 or 6 mm were used 
for most measurement with fluorinert as pressure transmitter. 
Complementary measurements were made with an ILL He-gas pressure cell. 
On D15, the pressure was determined from the lattice spacing of a NaCl crystal mounted 
inside the pressure cells.

The observed temperature dependence of the magnetic intensity ($\propto m^2$) 
of \urs\  is shown in Fig.\ \ref{MvsTpress}  for different pressures. 
The increase of the temperature slope, $dm^2(T)/dT$, with pressure near $T_M$ indicates that the phase 
transition is first order.  This behavior is opposite to that expected for a pressure gradient 
if the transition is second order, 
and is thus an intrinsic effect. 
Fitting the expression 
$m^2 = m_0^2 [1-(T/T_M)^\alpha]$  \cite{fak} leads to $\alpha\!\leq$2.5 for 
pressures below 5 kbar, 
while a much larger value, $\alpha\!\sim$8.5, is obtained at higher pressures,
suggestive of a first order transition. 
The phase diagram presented in Fig.\ \ref{phasediagram} is constructed from 
the data shown in Fig.\  \ref{MvsTpress},
where $T_M$ is taken as the {\it midpoint} of the transition, which is the correct estimate 
for a first-order transition in the presence of pressure gradients. 
The presence of pressure gradients, 
usually estimated to $\Delta p/p\sim5$\% for the used clamp cells, 
 is clear from the strong intensity observed at 4.5 kbar 
using a clamp cell (Fig.\ \ref{MvsTpress}),
since measurements in a He-gas pressure cell at $p=4.45 \pm 0.15$ kbar show no (large)
ordered moment on D15, where the detection limit is  0.06 $\mu_B$. 
We also note that the magnetic intensity saturates at a value of 0.33 $\mu_B$
for pressures $p\!\geq\!6.4$ kbar. 
The critical pressure for the onset of the large moment is estimated to $p_M=4.9 \pm 0.2$ kbar.  
As for the doped Rh sample (see inset of Fig.\ \ref{twopercent}), no hystereris was observed in $m^2(T)$. 
However, the vertical slope of $dT/dp$ in Fig.\ \ref{phasediagram} is a strong evidence of a first-order transition. 
The absence of magnetic scattering at the (0,0,2$l$+1) reflections shows that 
the ordered moments are along the $c$-axis at all pressures, 
i.e. the SMAF and LMAF phases have the same AFM structure.
We finally note that while the AFM Bragg peaks are resolution limited 
in the LMAF phase \cite{amiprl,id20}, 
they have a finite correlation length of a few hundred \AA\
in the SMAF phase \cite{broholm,fak,amiprl,id20}. 

The longitudinally polarized magnetic excitations were studied under pressure on IN22 
at the two wave vectors {\bf Q}$_1$=(1,0,0) and {\bf Q}$_2$=(1.4,0,0), 
which correspond to the minima in the $p$=0 dispersion curve of \URuSi\ \cite{broholm}.  
Spectra for energy transfers up to 7 meV were performed with
$k_f$=1.55 \AA$^{-1}$ 
and a 5-cm long Be filter in the scattered beam to reduce 
higher-order contamination; 
the energy resolution was 0.3 meV.  
Complementary spectra for energy 
transfers up to 16 meV were performed with
$k_f$=2.662 \AA$^{-1}$ and
a 4-cm long pyrolytic graphite filter in the scattered beam;
the energy resolution was 0.9 meV.

\begin{figure} 
\includegraphics[width=.8\columnwidth,clip]{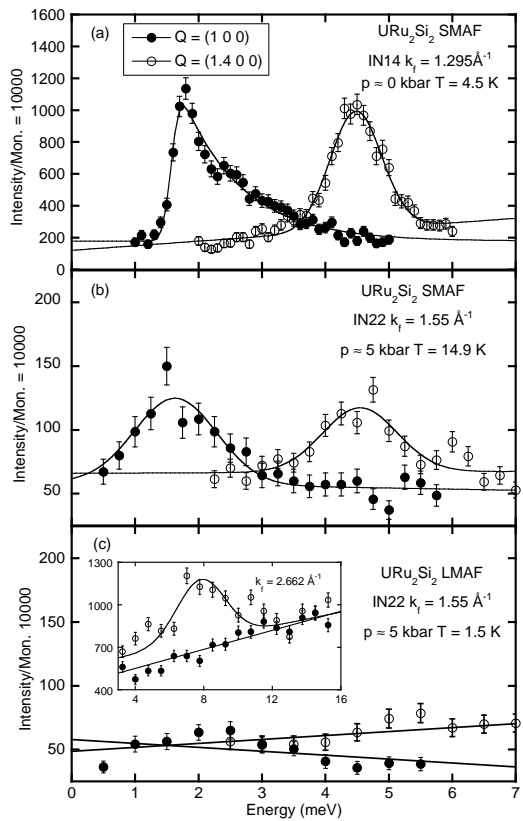}
\caption{Intensity {\it vs.} energy transfer for {\bf Q}$_1$=(1,0,0) and {\bf Q}$_2$=(1.4,0,0).  
(a) SMAF phase at $p=$0 and $T\!<\!5$ K shows 
two well-defined excitations at {\bf Q}$_1$ and {\bf Q}$_2$. 
(b) SMAF phase at  $p$=5 kbar and $T\!>\!7$ K also shows two well-defined excitations 
(the apparent broadening of the peaks is due to a coarser resolution and a higher $T$ for this measurement).  
(c) LMAF phase at $p$=5 kbar and $T$=1.5 K shows the absence of the  {\bf Q}$_1$ excitation 
and a strongly increased energy of the  {\bf Q}$_2$ excitation (inset). }
\label{excitations}
\end{figure}

The measurements show pronounced differences in the excitation spectra between 
the LMAF and SMAF phase.  In the SMAF phase, $T_M<T<T_m$, the two excitations 
at a finite pressures of 5.0 kbar are very similar to those at 
zero pressure, with the same gap energies of $\Delta_1$=1.6 and 
$\Delta_2$=4.5 meV at {\bf Q}$_1$ and {\bf Q}$_2$, respectively [see Figs.\ \ref{excitations}(a-b)].  
In the LMAF phase, however, the excitation at {\bf Q}$_1$ is absent [see Fig.\ \ref{excitations}(c)]. 
There is no clear feature at this wave vector 
in the measured spectra for energy transfers 
between 1 and 16 meV, in agreement with Ref. \cite{metoki}. 
This suggests that the {\bf Q}$_1$ excitation is closely related to $\psi$, 
which is small in the LMAF phase. 
The excitation at {\bf Q}$_2$, 
on the other hand, is readily seen. Its energy depends on pressure: we 
found $\Delta_2$=7.7, 8.6, and 9.2 meV at pressures of 5, 9, and 10.5 kbar, 
respectively. 

Our neutron scattering measurements on the 2\% Rh-doped sample reveal 
that the excitations (not shown) are very similar to the pure compound under pressure. 
In the SMAF phase, the excitations at both {\bf Q}$_1$ and {\bf Q}$_2$ are present.  
In the LMAF phase, the excitation at  {\bf Q}$_1$ is absent, as in \urs\ under pressure, 
while the excitation at  {\bf Q}$_2$ is still present. 

We will now discuss our results in the context of other work. 
Recent thermal-expansion measurements 
also observe a first-order transition under pressure \cite{sato}. 
The authors construct a phase diagram where the first-order transition line 
appears to join the second-order line around a pressure of 11 kbar. 
This would imply that the coefficient $\gamma$ in Eq.\ (\ref{landau}) is zero, 
in clear contradiction with high-field measurements \cite{FRL}.  
Our data, cf. Fig.\ \ref{phasediagram}, clearly show that the first and second order lines 
are nearly parallel at these pressures, 
and the most likely scenario is that the first-order line has an endpoint (cf. right panel of Fig.\ \ref{zhito}), 
and hence $\gamma\!\neq\!0$. 

NMR \cite{matsuda} and $\mu$SR \cite{amiuSR} measurements  have also observed
a large moment under pressure, i.e. in the LMAF phase. 
The coexistence of an AFM signal with a ``paramagnetic'' signal in the NMR measurements under pressure
and a magnetic field of a few Tesla on an aligned but crushed powder sample
by Matsuda {\it et al.}  \cite{matsuda} was interpreted in terms of a spatially inhomogenous AFM ordering. 
There are several arguments against such a scenario. 
First, it would be a curious coincident if the AFM and the hidden order would have 
the same transition temperature in a phase-separated scenario, 
since $m$ and $\psi$ are not coupled as they occur in different parts of the sample.
Secondly, one would not expect the {\bf Q}$_1$ excitation to disappear at a pressure 
where the two phases still coexist, 
just a smooth intensity change as the population of the two phases changes with pressure.
Thirdly, a phase-separated scenario is inconsistent with a number of other experiments, 
such as the recent de Haas-van Alphen measurements under pressure  \cite{nakashima}
and the NMR measurements by Bernal {\it et al.}  \cite{bernal}. 
Since the temperature dependence observed in Ref.\ \cite{matsuda} 
is suggestive of either a large pressure gradient or strong internal strains, 
it would be important to repeat $^{29}$Si-NMR measurements under pressure
on a well-characterized single crystal. 
We recall here that the magnetism of \urs\ is very sensitive to the sample quality \cite{fak}. 
Indeed, the neutron scattering measurement under pressure 
on an annealed but otherwise uncharacterized crystal (closely related to that of Ref.\ \cite{matsuda})
presented in Ref.\  \cite{amiprl} 
are partly inconsistent with 
our neutron scattering experiments,
the thermal-expansion measurements of Ref.\ \cite{sato}, 
and their own $\mu$SR work \cite{amiuSR}. 

In conclusion, both the $p$-$T$ phase diagram (this work) 
and the magnetic-field dependence of the ordered moment  \cite{FRL} in \urs\ 
are in excellent agreement with a scenario 
where a hidden OP is linearly coupled to the ordered moment 
and hence breaks time-reversal symmetry.

We acknowledge discussions with M.B. Walker and I. Fomin
and experimental support from A.A. Menovsky, A.D. Huxley, E. Ressouche,  
J.L. Laborier, F. Fauth, L. Paolasini, A. Bombardi, F. Yakhou and N. Pyka. We thank G. Motoyama and N.K. Sato for sending us complementary thermal expansion data.


\begin{thebibliography}{44}
\bibitem{calcp} The moment was estimated from the AFM energy in molecular field theory: 
$(m_0/m_{sat})^2\!=\!(2/3 k_B T_N)\int_{0}^{T_N}C_{mag}(T)dT$, 
with $m_{sat} = 3.62 \mu_B$.
\end{thebibliography}
\end{document}